# Performance Impact of Lock-Free Algorithms on Multicore Communication APIs

K. Eric Harper and Thijmen de Gooijer

Industrial Software Systems

ABB Corporate Research

eric.e.harper@us.abb.com, thijmen.de-gooijer@se.abb.com

*Abstract:* - Data race conditions in multi-tasking software applications are prevented by serializing access to shared memory resources, ensuring data consistency and deterministic behavior. Traditionally tasks acquire and release locks to synchronize operations on shared memory. Unfortunately, lock management can add significant processing overhead especially for multicore deployments where tasks on different cores convoy in queues waiting to acquire a lock. Implementing more than one lock introduces the risk of deadlock and using spinlocks constrains which cores a task can run on. The better alternative is to eliminate locks and validate that real-time properties are met, which is not directly considered in many embedded applications. Removing the locks is non-trivial and packaging lock-free algorithms for developers reduces the possibility of concurrency defects. This paper details how a multicore communication API implementation is enhanced to support lock-free messaging and the impact this has on data exchange latency between tasks. Throughput and latency are compared on Windows and Linux between lock-based and lock-free implementations for data exchange of messages, packets, and scalars. A model of the lock-free exchange predicts performance at the system architecture level and provides a stop criterion for the refactoring. The results show that migration from single to multicore hardware architectures degrades lock-based performance, and increases lock-free performance.

## 1 Introduction

The world is full of embedded devices whose software has been developed with the assumption that a single processor executes its functions. This fundamental constraint was based on the available low cost computer hardware at the time and permeated the design decisions related to resource trade-offs between computation, I/O and storage. Moore's Law [1] predicted the number of transistors on integrated circuits doubles approximately every two years. The densities of the transistors on a chip increased and the corresponding shorter distances between semiconductor devices reduced data propagation delays. Coupled with innovations in micro architectures, the available computing power measured by number of instructions performed per second (clock rate) increased by orders of magnitude over the past decades taking away the need to parallelize software designs for all common applications.

Moore's Law continues to hold true, but the power density limits [2] when using gigahertz (GHz) clock rates create an unanticipated roadblock for increasing single processor performance. The heat generated by an integrated circuit at such high clock rates cannot be passively dissipated and can damage the chip silicon. As a workaround the transistors are reorganized into multiple processors or cores on a single chip, where each processor has a lower clock rate but the combined parallel processing power is greater than previously achieved. These multicore processors are the available low cost computer hardware for the foreseeable future. Unfortunately the original assumption of single processor makes it difficult to take advantage of this commodity hardware without refactoring the software designs.

A study [3] of general purpose computing examples showed that multicore hardware can be leveraged to increase computational performance, but the parallelization strategy is important and cannot be delivered simply by recompiling single threaded source code or changing operating systems (OS). Multi-threading may be used to create parallelism but managing the threads is tedious and error prone. Using a concurrency runtime is a more reliable approach. Finally, Pankratius et al. suggest that shared memory is a much better programming model to achieve enhanced computing performance compared to message passing between HPC (High Performance Cluster) distributed memory nodes [3].

The strategy focus from Pankratius et al. holds true for embedded computing, a domain in which parallelizing sequential software is a challenging task [4]. Due to the difficulty of detecting and resolving concurrency issues in embedded designs, a structured approach using design patterns to gradually parallelize the design was applied in several case studies. The recommended steps [4] are to find concurrency in the application, look for architectural and structural patterns, and then decompose the logic into algorithmic and implementation strategy patterns. In all cases the software has to be refactored to take advantage of the parallel computing hardware.

There are two types of parallel programming models: data parallelism and task parallelism [5]. In data parallelism the same instruction is performed repeatedly and simultaneously on different data. In task parallelism, the execution of different work requests is distributed across multiple computing nodes. The key issues for task parallelism are synchronization and atomicity. Both task parallelism studies [3][4] indicate that data exchange latency between parallel tasks can be a bottleneck. In many cases this issue is left to the OS or a communication framework to resolve.

This paper shows how shared memory data exchange performance for embedded applications that are migrated from single core to multicore processor architectures can be improved up to twenty-five times by a lock-free design. We assume shared memory architecture on a single device, but with enough abstraction to make the embedded applications portable across different preemptive symmetric multi-processing (SMP) operating systems. For this paper only one-way first-in first out (FIFO) data exchange in a single address space is considered, but in future work we plan to report how we extend our work to other types of exchange and across more than one address space.

Our contributions are motivated by recent work replacing the primary locks in operating system kernels with fine-grained locks that allow more than one task to enter the kernel at a time [8][9]. This approach is necessary to enable parallelism between operating system tasks as the majority of their instructions are executed in kernel mode. The perspective for optimizing the locks can also be applied to application user mode instructions running on multicore computer hardware, and the best solution is to remove the locks entirely.

The rest of this paper describes a concurrency runtime with its APIs (Application Programming Interface) designed specifically for multicore (Section 2). We summarize the steps (Section 3) taken to refactor the runtime removing the shared memory access locks, along with the algorithms needed to implement lock-free data exchange. With stress tests (Section 4) designed to saturate the memory bus, we investigate how the Multicore Communications API (MCAPI, [21]) baseline and refactored implementations perform on single and multiple cores. Modeling at the system architecture level (Section 5) predicts lock-free performance and provides stop criteria for the refactoring. The results data (Section 6) indicate that lock-based I/O performance degrades on the Microsoft Windows® and Fedora Linux with real-time extensions operating systems as a result of the multicore migration. Removing the lock invocations provides a 25x speedup for multicore deployments on Linux. This represents a theoretical limit for shared memory I/O performance and techniques are suggested to achieve similar levels with more complex communication patterns on a single device. Finally, future work is proposed that can expand on our understanding of the impact of lock-free algorithms for multicore communication along with our Conclusions (Section 7).

## 2   Background and Related Work

Deployed computer hardware has many variations and changes faster than software [23]; the applications optimized for specific hardware become obsolete over time. Therefore, the programming interfaces and structures should be hardware and OS neutral. A good example of this approach is computer graphics rendering software libraries, where the hardware technology has evolved from simple memory devices to configurable units, and now to fully parallel processors [24]. There are two three-dimensional (3D) rendering APIs: 1)

DirectX (only available on Microsoft Windows OS) and 2) OpenGL[1]. Each has maintained and extended their interfaces for decades as the hardware performance increased by orders of magnitude.

Ideally, the applications would not need to be revised as the runtimes are adapted to hardware innovations in the future. For task communications this requires a neutral specification like MCAPI suitable for multicore systems. The first MCAPI document was released in 2008 and updated in 2011. This API is part of a broader strategy by the Multicore Association to define and promote industry standards that find common ground among proprietary approaches for multicore and multiprocessing software development on resource-constrained architectures. MCAPI is attractive for our work because it is targeted toward embedded deployments.

Work by the Multicore Association on the MCAPI specification lead to a roadmap that includes the Multicore Resource Management API (MRAPI) and Multicore Task Management API (MTAPI). In addition the association has working groups looking at programming practices, tools infrastructure and multicore virtualization. The MRAPI specification was released in 2010 to support MCAPI 2.0, and the MTAPI specification is in draft review hopefully to be released in 2013.

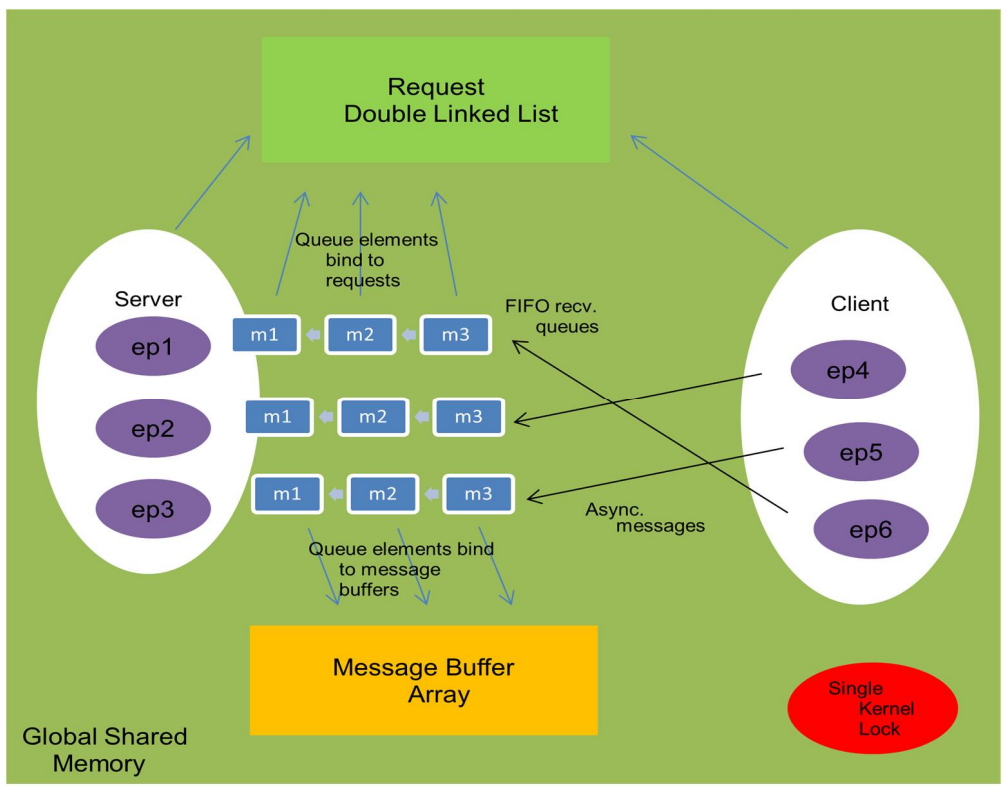

**Figure 1. MCAPI Reference Design**

The Multicore Communications API is responsible for synchronization and data movement between cores, both with blocking and non-blocking semantics. It is this abstraction that makes it possible to build applications without locks because all the data exchange is guaranteed to be performed without corruption. A task built on MCAPI can be programmed as if it were single threaded in relation to the tasks it communicates with, using messages for synchronization. MRAPI provides a separate API that abstracts OS resources and helps with external I/O communication.

---

[1] http://www.opengl.org

The MCAPI reference design is shown in Figure 1 above. MCAPI is built on top of MRAPI and implements platform-independent shared memory inter-process communication (IPC). There are three types of communication formats:

1) Messages - connection-less with priority-based FIFO delivery between ad-hoc endpoints,
2) Packets - connect-oriented delivery over established FIFO channels where the send buffer is provided by the caller and the receive buffer allocated from an MCAPI pool, and
3) Scalars - connection-oriented delivery over established FIFO channels for 8, 16, 32, and 64-bit values.

The data exchange structures, metadata and buffers are organized in a single shared memory partition. A user-mode reader/writer lock controls access to the partition and a single OS kernel lock guards changes to the reader/writer lock. Effectively, all write access to the global shared memory is serialized and the readers are blocked if a write is in progress. This shared memory database enables data exchange between the tasks and processes on a single device.

The client producer endpoints (ep) insert messages into the server consumer endpoint FIFO receive queues, where each queue entry (m) is bound to a reusable message buffer. Request objects managed in a double linked list are used to track asynchronous send and receive operations, allowing a separate task to complete an operation started by the originator. Consumers can reserve the queue elements that will be filled later by the producers.

The Multicore Resource Management API is responsible for memory management, user-mode synchronization and resources objects within a single task. MRAPI includes an underlying portability layer than exposes Unix System V Release 4 (SysVR4) type access to semaphores and shared memory. The resource structures and metadata are organized in a single shared memory partition guarded by a single OS kernel lock. User-mode mutexes, semaphores and reader/writer locks are built on top of this base. Resources are owned by nodes organized in domains, where the tasks can be mapped to the appropriate operating system and hardware resources, e.g. processes and threads. MRAPI abstracts connections to distributed memory and direct memory access (DMA) using the concept of remote memory that requires a custom implementation. Finally metadata management, including filtered resource trees and change triggered actions, is provided.

Our primary hypothesis is that removing the synchronization locks from software systems decreases I/O latency and increases throughput performance, and that this removal can be accomplished without compromising the data exchange reliability or integrity. There has been active research for operating systems regarding the impact of locks on multiprocessor systems starting in the 1990's. Given the assumption of kernel locks, Zahorjan [10] showed how the choice of task scheduling discipline could reduce the amount of spinning necessary. Karlin [11] identified that the cost of blocking one thread and activating another can be a substantial part of the program execution time. Mellor-Crummey [12] suggested that it is possible to construct busy wait synchronization algorithms that induce no memory or interconnect contention.

Michael [13] was one of the first to compare non-blocking to lock-based queues and found the lock-free approach consistently outperformed the best known alternatives. Nikolopoulos [14] compared spin locks, queues and barriers and showed the lock-free implementations performed better by an order of magnitude. Tsigas [15] hypothesized that software synchronization mechanisms result in poor performance because they produce convoy effects in multiprogramming environments, and performed experiments on the same system as Nikolopoulos to demonstrate that non-blocking synchronization performs as well, and often better than the respective blocking synchronization.

The previous works that significantly informed the approach of this paper are two contributions related to event and state message communication. Kopetz [16] summarizes the lock-free algorithm for a non-blocking write protocol (NBW) and formalizes the real-time properties of the solution. Kim [17] extends use of NBW to design the lock-free algorithm for a FIFO non-blocking buffer (NBB) and how it can be composed to support complex communication patterns including publish / subscribe and broadcast connections.

Lock-free algorithms guarantee a program's progress in a finite time period. Wait freedom is an alternative and has a stronger property: each thread is ensured progress. Kogan [18] surveys the contributions for wait-free algorithms and finds that they are inefficient and hard to design because of the work-stealing approach and making sure a request is only executed once. This is an area for future investigation if it is determined the lock-free techniques result in some data exchange participants being starved for access.

The defacto standard for parallel computing communication is the Message Passing Interface (MPI) [19]. The API was developed especially for distributed memory HPC but has been adapted for shared memory. The objections to MPI have been its large memory footprint and lack of support for dynamic processes. OpenMP [20] is a similar API with incrementally better performance than MPI, but with the goal of supporting distributed HPC.

The Multicore Association was formed in 2005 to go the other direction, focusing on communication at the hardware level rather than between computers. Holt writes [21] about the MCAPI standard and recommends it for small footprint, highly efficient intercore and interchip communications. More recently Gray [22] combined MCAPI with a layered architecture to support more general purpose programming rather than having a specific hardware focus.

## 3   Contributions

The Multicore Association focus is at the computer hardware level, but the APIs are applicable wherever concurrency is needed. For example, an embedded application might be first demonstrated on a general purpose platform such as Microsoft Windows with its rich development and debugging environment, and then ported and validated on a real-time operating system (RTOS). Second, there are advantages to running the embedded applications on non-RTOS platforms to configure and experiment with industrial deployments rather than trying to emulate the RTOS environment.

The MRAPI reference implementation has a number of innovations. First, the SysVR4 foundation is common to most embedded platforms and provides a workable portability abstraction. Second, managing the resources in a shared memory partition facilitates consistency across the real-time processes and makes it possible for the resource configuration to be initialized from a disk image at startup. The user-mode locks are lighter weight than the OS kernel locks and do not require a context switch to engage. Finally, organizing nodes in domains has security benefits where the resource access across domains might require authentication to prevent malicious attacks.

We ported the MRAPI reference implementation to run on Microsoft Windows Server 2008 R2, including support for OS handles, semaphores, shared memory and event-based signals. The other fundamental gap in MRAPI, compared to other concurrency runtimes, is the lack of atomic CPU (Central Processing Unit) instructions which can act as very lightweight locks. Operating systems implement access to the underlying processor atomic machine instructions in a non-portable fashion, and so cross-platform access functions were added to the portability layer, including barrier, compare-and-swap and bit operations. This was straightforward for tasks running in the same address space, but more difficult for RTP (Real-Time Process) synchronization running on non-Windows platforms. Finally, MRAPI was extended to provide platform independent explicit context switching and timed delay.

There are a number of bottlenecks in the original MCAPI implementation, but the most expensive are MRAPI lock invocations for every asynchronous request or data exchange. These locks were disabled to reduce data exchange latency. Eliminating locks from embedded applications is non-trivial and very risky because all the existing software interactions are built with the assumption that concurrent access to shared data is strictly prevented. The best practice [4] is to proceed carefully; make incremental changes to a stable baseline and regularly validate whether the revisions are successful. The guard on the global shared memory, shown as the red oval at the bottom left hand corner of Figure 1, can only be disabled when all the underlying data structures are immune to concurrent access.

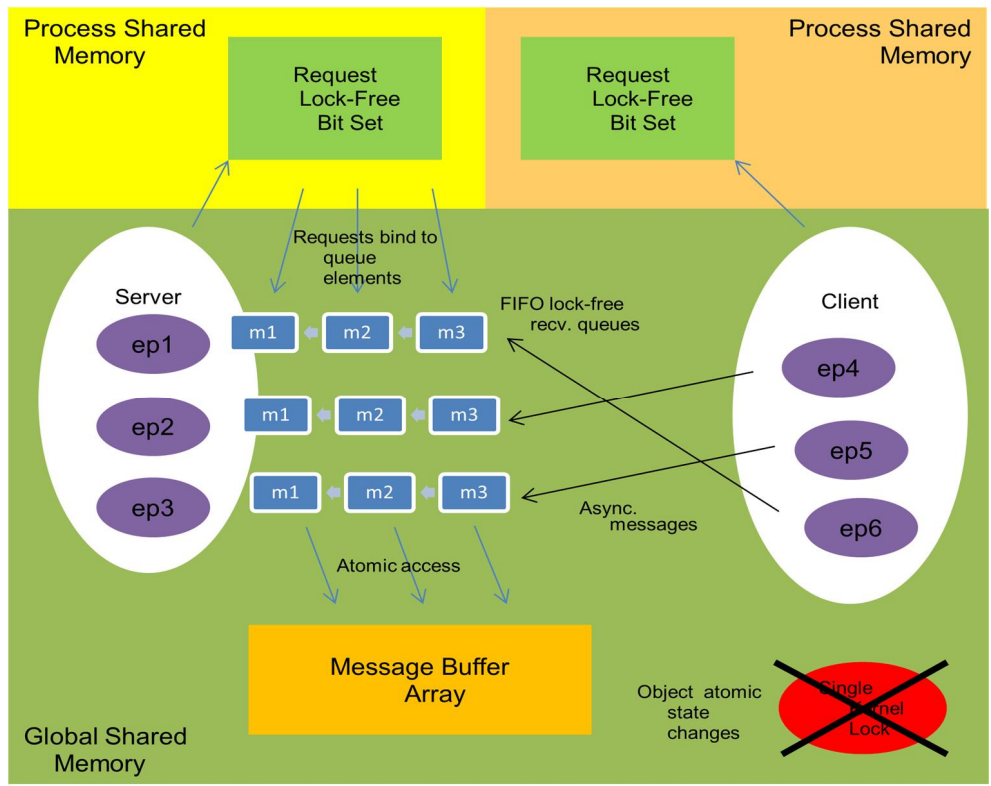

**Figure 2. Lock-free MCAPI Design**

MCAPI refactoring proceeded in four steps, with the end result shown in Figure 2 above.

1) Move the request objects from global to process shared memory and convert the request double linked list to a lock-free double linked list [25],
2) Convert the endpoint FIFO queues to lock-free FIFO queues [17],
3) Replace the lock-free request double linked list with a lock-free bit set (because lock-free double linked lists are not feasible [26]), and
4) Ensure all runtime access to communication metadata is done with atomic operations to allow reliable node run-up and rundown.

A key requirement for most applications is data consistency: one task should not corrupt the data another task is using. A number of enabling conventions are already in place, especially for embedded systems. First, the smallest random access memory (RAM) mutation is a byte. Access to system main memory depends on the underlying CPU hardware and specified coherency protocol. For example, the PowerPC [27] reads and writes to memory in bytes [28], where each byte access is atomic and multiple bytes can be read or written in any order. With multiple processor cores this means copying any data type larger than a byte must be explicitly protected when there is contention for the location between cores.

The RTOS development tool chain for the embedded system can provide these explicit atomic operations. Based on the authors' prior experience, the source code compiler used to build embedded VxWorks[2] images generates atomic CPU instructions for four byte memory writes by default. This makes programmatic atomic operation syntax unnecessary. Copying data types larger than four bytes needs explicit synchronization not provided by the tool chain to ensure reliability. For the PowerPC, atomic operations on the same memory location are executed in program order without need for a memory barrier. Data consistency across address spaces and for non-atomic types requires explicit programming instructions.

---

[2] http://www.windriver.com

There are two classes of message communication [17]: state and event. The non-blocking write (NBW) protocol [16] is used for lock-free state messaging and also serves as the pattern for lock-free event messaging. From Kopetz, et. al., lock-free messaging has the following properties:

1) Safety – if a read operation completes successfully, guarantee that it has read an uncorrupted version,
2) Timeliness – tasks containing the read operations must complete their execution before deadline, and
3) Non-blocking – the writer cannot be blocked by readers.

The order that state messages are exchanged is indeterminate; they simply deliver the current value. On the other hand, event message order is preserved. The algorithms are implemented with atomic counters. For state messages there is a single atomic counter, with initial value set to zero. When the counter overflows it is set back to zero. The approach is similar to optimistic locking in databases [29]. Each time the writer has a new message, it first increments the counter, writes the message in the next available array buffer (typically associated with the counter value), and then increments the counter again. A reader grabs the value of the counter, reads the message in the associated array buffer, and then checks to see if the message contents were corrupted by a concurrent write. The collision test is performed by comparing the counter value before and after the read. If the message was overwritten during the read, the reader attempts to read again. The more array buffers there are, the less likely a collision will occur between reading and writing.

The Non-blocking Buffer (NBB) is used for lock-free event messaging. As described by Kim, et al., we use two atomic counters, one for the writer and one for the reader. They are both managed similar to the single state message counter. The underlying data structure is a circular ring buffer FIFO queue with one counter controlling synchronization for update and the other for acknowledge ensuring the writer and reader always access different slots in the ring buffer. The size of the NBB needs to accommodate message bursts.

| InsertItem | ReadItem |
| --- | --- |
| `BUFFER_FULL` – no room for new items, caller should yield processor and retry, perhaps after some delay | `BUFFER_EMPTY` – no pending items to read, caller should yield processor and retry, perhaps after some delay |
| `BUFFER_FULL_BUT_CONSUMER_READING` – no room for new items, caller should not yield processor; retry immediately a limited number of times with no delay | `BUFFER_EMPTY_BUT_PRODUCER_INSERTING` – no pending items to be read, caller should not yield processor; retry immediately a limited number of times with no delay |

**Table 1. Non-Blocking Buffer Operation Errors**

The two counters guard sections of the ring buffer: 1) the portion available to write, and 2) the portion available to read. Similar to NBW, each counter is incremented before an operation starts and then again after it completes. The basic operations are InsertItem and ReadItem. The operations succeed or return one of two errors as shown in Table 1 above.

The same design properties from NBW can be tested here:

1) Safety – guarantee of uncorrupted read holds true,
2) Timeliness – read operations either complete successfully with no delay, fail with a limited number of immediate retries (not compromising the deadline), or fail with an indication to attempt the read in the next cycle, meaning the application is responsible for timeliness, and
3) Non-blocking – write operations either complete successfully with no delay, or fail with a limited number of immediate retries (not blocking further processing) if the reader is stalled for any reason, which means the application is responsible for overall non-blocking.

These lock-free algorithms were refactored into the MCAPI implementation keeping the original MCAPI software design structure largely intact. One key change was to use finite state for all object state transitions similar to [30] and verify with atomic compare-and-swap that an object is in the expected state before changing to the next state. Debugging race conditions is difficult in high throughput, low latency designs. Due to the observer effect, the introduction of measurements or recording log output may cause changes in the runtime dynamics. Concurrency defects are either hidden or shifted to other locations in the code.

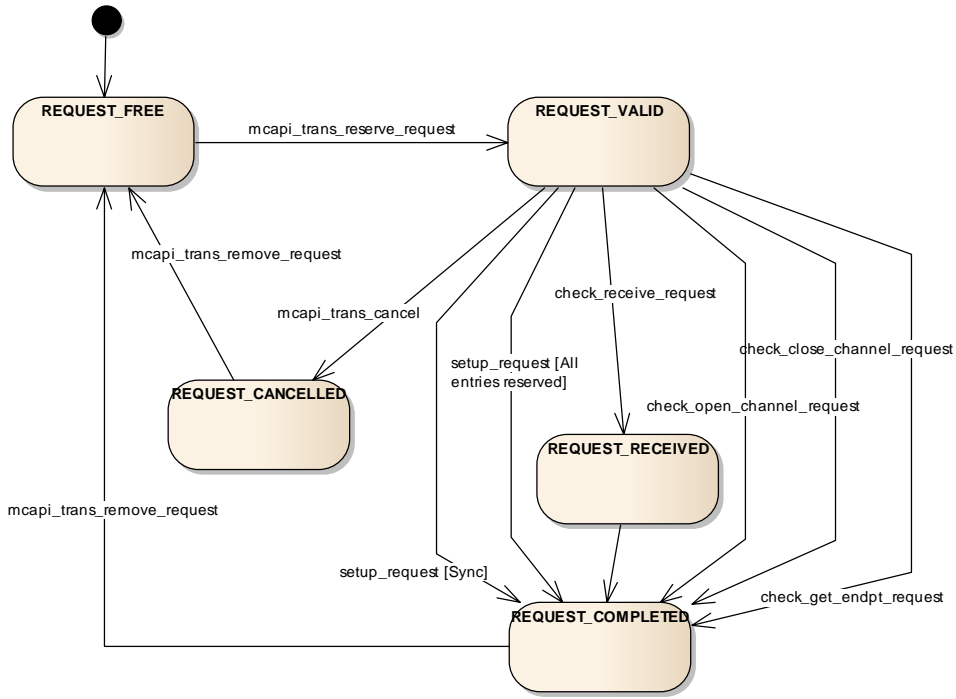

**Figure 3. MCAPI Request Transitions**

For example, request objects in the original implementation are marked with several boolean flags that indicate whether the request is valid, completed or cancelled. These flags were replaced with the state transition diagram shown in Figure 3 above. A request in the REQUEST_FREE state is available for any client in that address space to identify a pending asynchronous operation, e.g. opening a channel, sending a message, etc. Once the request is allocated its state changes to REQUEST_VALID.

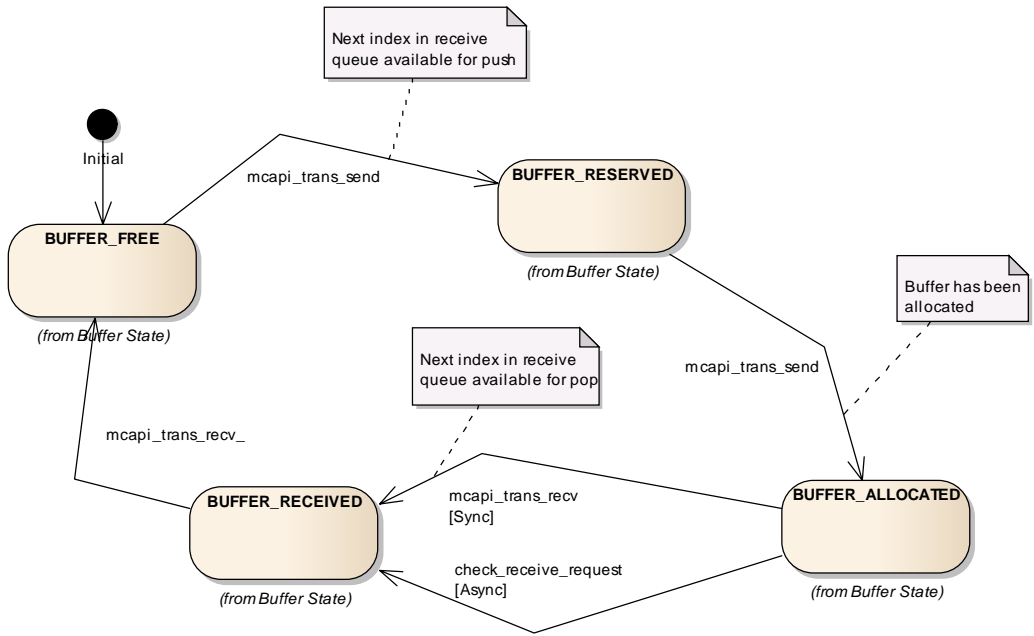

**Figure 4. MCAPI Queue Entry Transitions**

For all the operations other than asynchronous send, request completion changes the state to `REQUEST_COMPLETED`. For the exceptional send case, the request is marked as `REQUEST_RECEIVED` until the buffer can be confirmed received, and then the request state changes to `REQUEST_COMPLETED`. The request is then transitioned back to the available pool by changing its state to `REQUEST_FREE`. Cancelling a pending receive request (send requests always complete) changes the state to `REQUEST_CANCELLED`, and then `REQUEST_FREE` to make the cancelled request ID available for future operations.

Receive FIFO queue entries are marked in the original implementation with a boolean flag indicating if the entry is valid or not. This flag was replaced with the state transition diagram shown in Figure 4 above. A queue entry in the `BUFFER_FREE` state does not have a buffer associated with it. Once a queue entry is available to receive a message it transitions to the `BUFFER_RESERVED` state. This guards the entry from use by other clients until a free buffer can be linked to the entry, and then the state transitions to `BUFFER_ALLOCATED`. When that message is at the head of the receive queue it is marked as `BUFFER_RECEIVED` to keep other listeners on the same endpoint from trying to read its buffer. The queue entry returns to the `BUFFER_FREE` state when the receive operation is complete.

## 4 Concurrency Test Design and Environment

We managed the process of MCAPI lock-free refactoring using a test-driven development (TDD) approach [31]. A set of unit tests exercises the internal APIs that make up the runtime implementation. The external APIs (according to the Multicore Association specifications) are thin wrappers over the internal functions. Each internal function is called in turn with invalid and valid parameters using a single thread. Internal data structures are exposed with white box techniques to validate for expected pre- and post-conditions implemented as assertions [32]. The unit tests act as a safety net so as the implementation is revised, the test execution rapidly reveals any regressions.

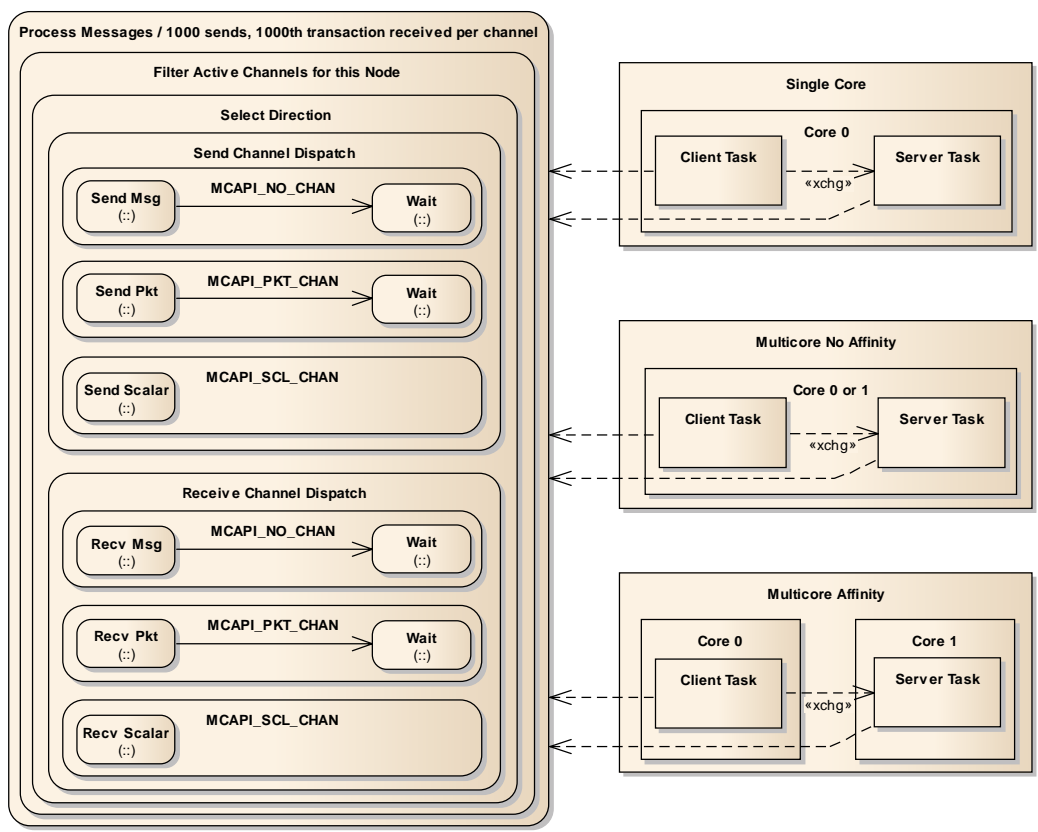

**Figure 5. MCAPI Stress Test Deployments**

Having these tests increases confidence for taking risk in major refactoring experiments. For example, first the lock-based MCAPI reference implementation was built and verified using an initial version of the unit tests. Then the lock invocations were incrementally removed as the unit tests verified continuously there were no regressions. Finally, the unit tests were updated to reflect and confirm the internal white box changes made to enhance the lock-free implementation.

The real test of the concurrency runtime is when the algorithms are exposed to high frequency requests. This stress is applied within a single process by launching multiple threads that act as the clients and servers communicating without any explicit delays between the requests. The communication paths and directions are configured by a declarative message topology designed by the authors, and each operation is marked with a monotonically increasing transaction ID so it can be tracked to completion.

A single routine was designed to run in each of the client and server nodes, one thread per node. The schematic in Figure 5 above shows the processing routine as a set of nested dispatches inside of a loop that iterates repeatedly over the configured channels. All the communication channels are set up before the loop starts, and they are run down and released when the processing loop completes. The loop exits when the following is true:

1) Each active channel with a send endpoint for the running node has transmitted one thousand messages with transaction IDs 1 through 1000, and
2) Each active channel with a receive endpoint for the running node has accepted a message with transaction ID 1000.

Consider a simple example. Two nodes are configured with a single channel between them, where the first node (node1) is responsible for the send endpoint and the second node (node2) is responsible for the receive endpoint. Each node starts its processing routine simultaneously on a different thread. Node2 immediately posts an asynchronous read request according to the channel type on the receive endpoint and loops, tracking the request to completion. For the messages types other than scalar, Wait is invoked with an immediate timeout, followed by yielding control to other threads. Scalar messages either succeed or fail immediately. Once a read completes the transaction ID is verified to be in sequence. In parallel, Node1 immediately sends an asynchronous message according to the channel type with transaction ID 1 and tracks the request to completion. The Wait is invoked with an immediate timeout, followed by yielding control to the other threads.

For more than one channel the processing iterates over the configuration in a round robin fashion. The stress tests are run:

1) With all the threads set for CPU affinity to one core,
2) The threads allowed to run on any core (no affinity) and
3) With the threads set for CPU affinity each assigned across the available cores.

The sender typically executes without interruption until the receive queue is filled, and then yields. Once the queue is full the receiver takes over and continues without blocking until the queue is empty. Then the receiver yields, allowing the sender to fill the queue again.

An eight-core server running CentOS SMP Linux 2.6.32-279.14.1.el6.x86_64 was used as the host for the experiments, with identically configured KVM (Kernel-based Virtual Machine) guests set up to run the targeted test operating systems. The host hardware specifications are dual socket quad core 2.5GHz Intel Xeon processors with E54202x6MB cache and 1333MHz FSB, with 667MHz 16GB random access main memory (RAM) and 1TB disk configured as a redundant array of independent disks (RAID5). The virtual machines are set up as 64-bit guests allocated each with four cores and 2GB RAM. The Windows VM is installed with Microsoft Windows Server 2008 Enterprise Service Pack 1 and the non-Windows VM with Red Hat Fedora 15 SMP Linux 2.6.43.8-1.fc15.x86_64 and real-time kernel extensions[3].

---

[3] http://ccrma.stanford.edu/planetccrma/software

# 5  Performance Model

Performance models can predict the system performance before implementation, and can be used to quickly study the design alternatives. Performance modeling may increase understanding of a system's performance problems and the model simulation results can help steer discussions with the system stakeholders [33]. The tests of API implementation alternatives show the speed-ups achieved in simple message structures and allow raw throughput measurement. However, we can neither decide whether the maximum performance has been reached, nor do these tests tell us anything about the performance of the overall system. We used a Queuing Petri-Net (QPN) performance model[4] to address these two issues. Technical details about our model and lessons learned will appear in another publication.

Due to the performance gap between CPU speed and memory access times [34], increasingly many cores share the same memory in multi-core SMP systems. Therefore we assume the shared memory resource will be the bottleneck for our lock-free message exchange rather than the computing capacity in the individual cores. After removing the bottleneck of the shared locks, the shared memory is the next one-lane bridge that message transactions will try to use concurrently. The QPN model has a single queue representing the shared memory and a limited, configurable number of cores. The model's Petri-net places and transitions represent the architecture of our system. Colored Petri-net tokens flow through the system and represent the expected communication workload between the components of the system.

UML (Unified Modeling Language) sequence diagrams for the lock-based and lock-free message exchange implementations were created using static analysis. The number of memory operations needed for sending and receiving a message were computed from these diagrams. We collected data about memory access times from various public benchmarks (e.g., [35]). How often the memory is accessed and how long one memory operation typically takes were encoded as the resource demands in the QPN model.

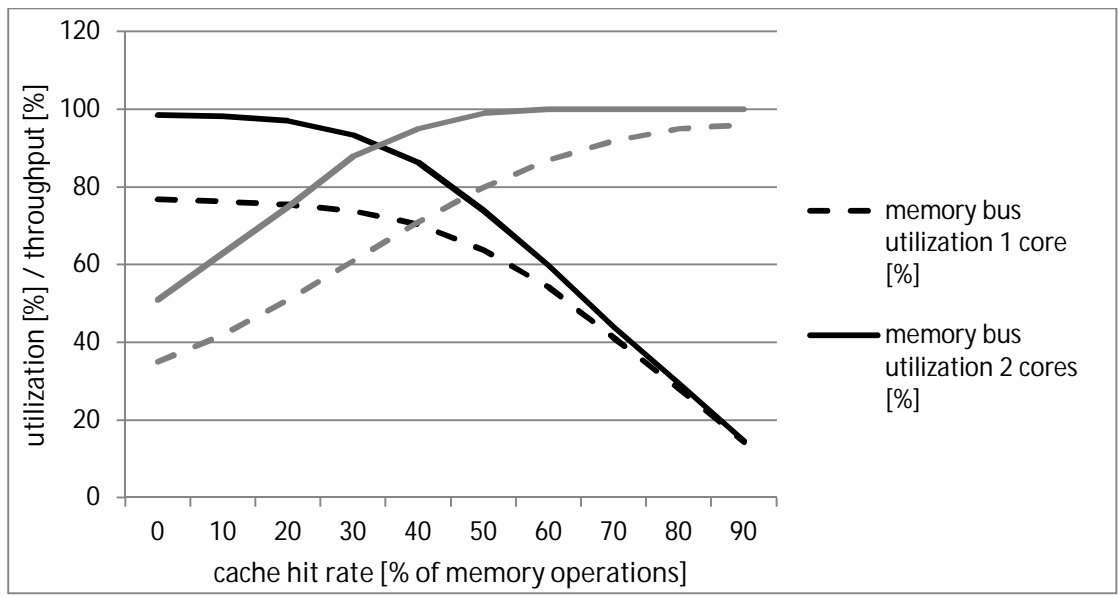

**Figure 6. QPN Model Simulation Results**

Having modeled the system behavior, structure and workload, we can simulate the model for lock-free performance while changing the parameters such as cache hit rate, number of cores, and task to core allocation. Obviously, the limited fidelity of the model means the results are not perfect. For example, the CPU time required for the message exchange was not considered, nor did we model any of the other workloads that exist on the system that might interfere with the message exchange. Furthermore, our model only considered main memory access and did not try to capture the complicated caching behavior of modern systems. We simply assumed a cache hit would not result in any demand on the main memory and would not require pre-fetching.

---

[4] We used the QPME tool (http://descartes.ipd.kit.edu/projects/qpme/).

A concise overview of some of the results is shown in Figure 6 above. It shows the memory bus utilization and the message throughput in percent for one of the system's message types. The dotted lines represent a single core model configuration, the solid lines a dual core configuration. The metrics are plotted as a function of cache hit rate.

Taking a more detailed look at Figure 6, we see that the workload cannot fully saturate the memory bus resource when only a single core is available (black dotted line). Regardless of the hit rate, we do not attain the target throughput rate but only about 95% for the displayed message type (dotted grey line). Note that we show only one message type in the graph, other workloads are excluded to improve legibility. Adding a second core raises the memory bus utilization as expected (black solid line), improves the throughput (solid grey line) but increases the contention for the memory bus. The target throughput is achieved only at relatively high cache hit rates. We can now ask what a realistic cache hit rate might be for the message exchange, but it is clear that we are pushing the system to its limits. Even at a ninety percent cache hit rate the memory bus utilization due to the message exchange is around fifteen percent without considering pre-fetch workloads, other applications, and interference.

Simulation of model configurations similar to our test environment and theoretical calculations based on the data gathered for the model gave us a theoretical maximum message throughput rate of 630,000 messages per second or one message every 0.63 microsecond. The minimum measured elapsed latency of the lock-free implementation on Linux is seven microseconds, an order of magnitude higher than the theoretical maximum. However, the theoretical maximum only considers the time required for the cache and memory transactions by the message exchange and excludes the time needed for execution in the CPU, the atomic instructions, and operating system tasks. Furthermore, the calculated maximum does not model the FIFO requirement that is implemented. In future work we will refine our implementation and study messaging models other than FIFO to seek throughputs closer to our computed maximum.

## 6   Results

The stress tests were used to exercise the MCAPI concurrency runtime for single one-way data exchange considering a test matrix with four different dimensions:

1) Windows vs. Linux deployment on the same hardware,
2) Single core vs. multicore resources,
3) Message type: message, packet, and scalar, and
4) FIFO lock-based vs. lock-free.

| Multicore lock-based MCAPI throughput speedup | Task | Affinity Task |
|---|---|---|
| *Windows* | | |
| Message | 0.74x | 0.74x |
| Packet | 0.67x | 0.68x |
| Scalar | 0.80x | 0.69x |
| *Linux* | | |
| Message | 0.23x | 0.22x |
| Packet | 0.22x | 0.21x |
| Scalar | 0.24x | 0.22x |

5)   **Table 2. Lock-Based MCAPI Multicore Penalty**

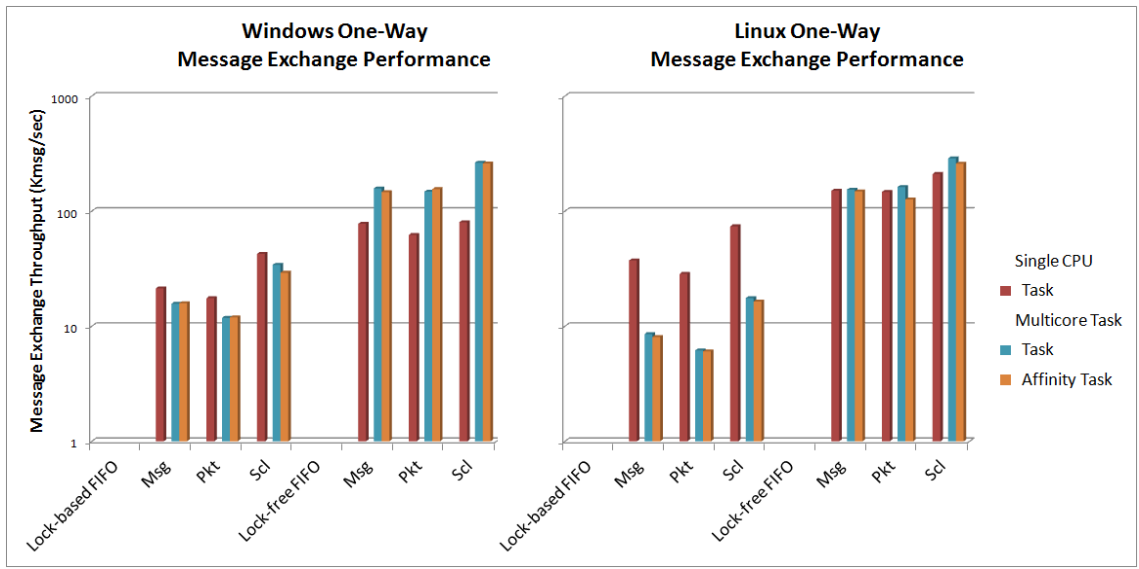

**Figure 7. MCAPI Data Exchange Throughput Performance**

Typical message and packet sizes are around twenty four bytes. The tests for the multicore scenarios were with the threads constrained to run on specific cores using CPU affinity vs. no affinity (i.e. dynamically allocated).

On both Windows and Linux, lock-based FIFO performs better on single core than multicore. The throughput speedups, defined to be:

```
Throughput speedup = (test throughput) / (original throughput)     (6-1)
```

are shown in Table 2 above. Using CPU affinity does not appear to make a significant difference and the penalty from migrating to multicore is more significant on Linux by at least a factor of three.

Our performance model (Section 5) suggests that the memory bus is saturated in the multicore tests, which means adding more channels would degrade the performance of each channel. The data exchange throughput performance for a single one-way channel as measured in thousands of messages per second is shown in Figure 7 above.

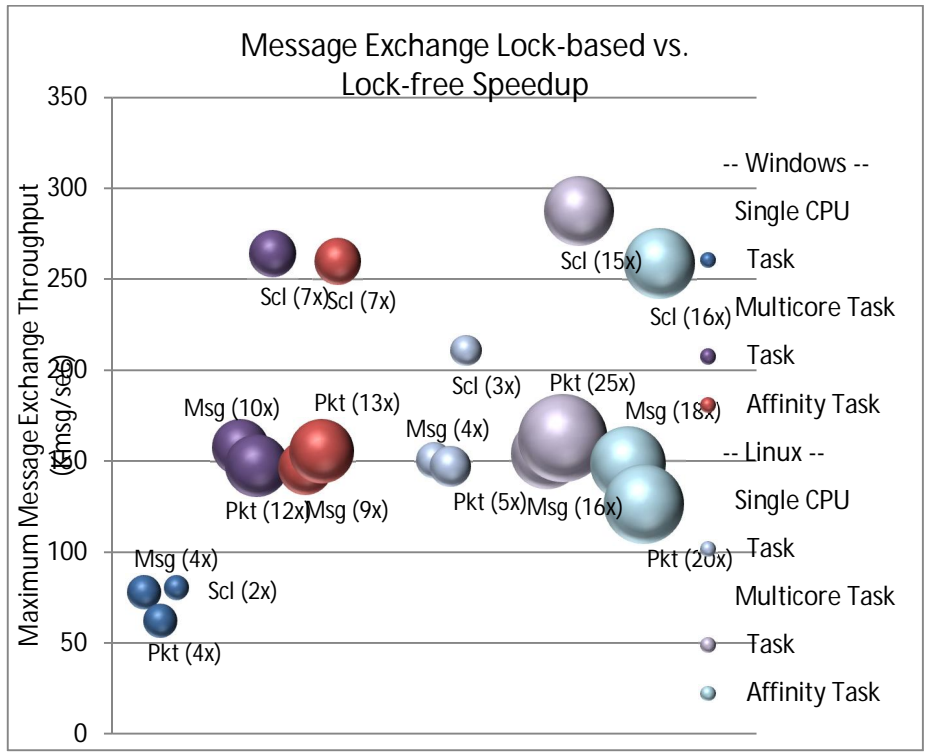

**Figure 8. Lock-free MCAPI Speedup**

Another way to visualize MCAPI data exchange performance is shown in Figure 8 above. The horizontal axis separates the test matrix (Windows vs. Linux, etc.) and the vertical axis is throughput performance measured in thousands of messages per second. The size of a bubble is based on latency speedup, defined to be

```
Latency speedup = (original latency) / (test latency)   (6-2)
```

and each bubble is positioned at the lock-free throughput measurement. The smallest bubble corresponds to about 2x speedup, and the largest bubble represents a speedup of 25x.

Larger bubbles indicate a bigger payoff that can be gained from investing development resources to make the change to lock-free FIFO. Changing from lock-based to lock-free on single core provides only incremental improvement. The most dramatic change comes from getting rid of the locks on multicore. Finally, on Linux using CPU affinity actually reduces the throughput performance.

The bandwidth capacity achieved in these tests is a worst case scenario because of the fine-grained nature of the data exchanges. The primary I/O bottleneck is the latency associated with transferring ownership of shared memory buffers from sender to receiver, independent of the size of the buffers. Combining multiple messages into a single packet buffer can increase the throughput by orders of magnitude more. On the other hand, fitting a message into a processor L1/L2 cache line and exchanging the data between two tasks running on the same core has the potential to dramatically decrease latency without need to transit the main memory data bus [36].

## 7   Conclusions

We have ported the MCAPI concurrency runtime to the Microsoft Windows technology stack and extended the MRAPI specification to provide first class portable access to atomic CPU operations. The primary shared memory I/O performance bottleneck has been removed by using lock-free techniques. Based on our measurements we conclude that lock-based algorithms for multicore communication using shared memory perform better on a single core compared to the same implementation running on multicore. On the other hand, lock-free data exchange has lower latency than lock-based techniques for both single and multicore deployments. Furthermore, the lock-free data exchange implementation enables applications to take advantage of the multicore hardware. For fine-grained messages, configuring CPU affinity and thereby restricting multicore tasks to run on specific cores does not give a significant throughput increase.

We have learned that migrating embedded applications to use multiple cores requires deep understanding of the design, either from previous development artifacts or through reverse engineering. Some traditional embedded implementations take advantage of single address space to share global data between tasks, depending on the natural serialization enforced by a single CPU to ensure the data is not corrupted. Based the authors' experience a better practice is to use interfaces, such as MCAPI, for data exchange between tasks. Traditional shared memory implementations use OS kernel locks to synchronize access. The presence of locks to guard data exchange hides a wealth of issues not just related to data exchange when initially migrating to multicore, and each of these must be dealt with in a methodical fashion. Removing the locks pays dividends with dramatic I/O performance increases.

In future work, we plan to enhance the MCAPI runtime to support state message data exchange policies and to enable MRAPI atomic operations across real-time processes. We expect to see a speed-up with the state message exchange policy, because it drops the FIFO requirement. Hopefully this and other refinements will bring measured performance closer to the theoretical maximum performance given by our model and calculations. Another way the current work could be extended is by counting the number of machine instructions in the implementation as a measure of the CPU time and adding this information to the performance model.